# Computational Design of the Rare-Earth Reduced Permanent Magnets

A. Kovacs[a], J. Fischbacher[a], M. Gusenbauer[a], H. Oezelt[a],
H. C. Herper[b], O. Yu. Vekilova[b], P. Nieves[c], S. Arapan[c,d], T. Schrefl[a,*]

[a] Department for Integrated Sensor Systems, Danube University Krems, Austria

[b] Department of Physics and Astronomy, Uppsala University, Uppsala, Sweden

[c] ICCRAM, University of Burgos, Burgos, Spain

[d] IT4Innovations, VSB-Technical University of Ostrava, Ostrava, Czech Republic

* Corresponding author. E-mail: thomas.schrefl@donau-uni.ac.at, Tel: +43 2622 23420 20, Fax: 43 2622 23420 99, Address: Donau-Universität Krems, Department für Integrierte Sensorsysteme, Viktor Kaplan Str. 2 E, 2700 Wiener Neustadt, Austria

Column of article (Research) Type of article (Article) Subject: <u>Rare earth permanent magnets</u>

**ABSTRACT** Multiscale simulation is a key research tool for the quest for new permanent magnets. Starting with first principles methods, a sequence of simulation methods can be applied to calculate the maximum possible coercive field and expected energy density product of a magnet made from a novel magnetic material composition. Fe-rich magnetic phases suitable for permanent magnets can be found by adaptive genetic algorithms. The intrinsic properties computed by ab initio simulations are used as input for micromagnetic simulations of the hysteresis properties of permanent magnets with realistic structure. Using machine learning techniques, the magnet's structure can be optimized so that the upper limits for coercivity and energy density product for a given phase can be estimated. Structure property relations of synthetic permanent magnets were computed for several candidate hard magnetic phases. The following pairs (coercive field (T), energy density product (kJ/m³)) were obtained for $Fe_3Sn_{0.75}Sb_{0.25}$: (0.49, 290), $L1_0$ FeNi: (1, 400), $CoFe_6Ta$: (0.87, 425), and MnAl: (0.53, 80).

## 1 Introduction

Permanent magnets are widely used in modern society. Important markets for permanent magnets [1] are wind power, hybrid and electric vehicles, electric bikes, air conditioning, acoustic transducer, and hard disk drives. With the growing demand for permanent magnets in environmental friendly transport and power generation [2] there is an ongoing quest to reduce the rare-earth content or use alternative rare-earth efficient or rare-earth free hard magnetic phases. Some of the considered hard magnetic phases may bridge the gap between ferrites and high performance NdFeB magnets [3].

In this work we present an overview on how the magnetic properties of a virtual magnet can be predicted starting from first principles. The materials modelling workflow in this paper is an example for traditional multiscale simulations with parameter passing. Several physical models are linked together in order to compute the hysteresis properties of permanent magnets: Genetic algorithms in combination with density functional theory guide the search for stable uniaxial ferromagnetic phases. This process may be assisted by mining materials databases. Then density functional theory is applied, in order to compute intrinsic magnetic properties such as spontaneous magnetization, magneto-crystalline anisotropy energy, and exchange integrals. The results feed into atomistic spin dynamics models for the computation of the magnetization, the anisotropy constant, and the exchange constant as function of temperature. These temperature dependent properties are then used as input for micromagnetic simulations. Numerical optimization tools help to tune the microstructure such that the coercive field or the energy density product is maximized for a given set of intrinsic magnetic properties.

In addition, we computed the reduction of coercivity owing to thermal fluctuations [4]. Analyzing the results of the micromagnetic simulations, we can identify how much different effects such as misorientation, demagnetizing fields, and thermal fluctuations reduce the coercive field with respect to the anisotropy field of the material.

The main focus of this paper is to predict the potential for various rare-earth free or rare-earth reduced permanent magnetic phases with respect to the expected extrinsic magnetic properties such as coercivity and energy density product. A sufficiently high coercive field and a sufficiently high energy density product are the key for the application of a new phase. These properties result from the interplay between the intrinsic magnetic properties of the magnet, the magnet's microstructure, and thermal fluctuations. Therefore, the main part of the paper will cover micromagnetic results for the hysteresis properties, which will be presented for well-known magnetic phases ($L1_0$ FeNi,

Nd$_{0.2}$,Zr$_{0.8}$Fe$_{10}$Si$_2$, Sm$_{0.7}$Zr$_{0.3}$Fe$_{10}$Si$_2$) as well as for phases predicted by genetic algorithms and density functional theory (Fe$_5$Ge, CoFe$_6$Ta). For some of the phases the intrinsic magnetic properties were computed by first principle simulations (Fe$_5$Ge, CoFe$_6$Ta, Fe$_3$Sn$_{0.75}$Sb$_{0.25}$) and atomistic spin dynamics (MnAl).

The results presented in this paper are centered on the micromagnetic computation of the expected performance of various hard magnetic phases. For details on the adaptive genetic algorithms for the search for new magnetic phases we refer the reader to recent articles applying the method to Fe$_3$Sn [5], CoFe$_2$P [6], and magnetic phases in the L1$_0$ structure [7]. First principle simulations of magnetic properties are review in [8]. An overview of essential micromagnetic techniques to compute the influence of microstructure on the coercivity and on the energy density product is given by Fischbacher et al. [9].

## 2 Methods

An adaptive genetic algorithm [7] in combination with the ab initio package VASP [10] was used to scan the phase space for Fe rich compounds that are non-cubic and stable. The magnetic properties were calculated with help of the full-potential linear muffin-tin orbital (FP-LMTO) method implemented in the RSPt code [8]. Synthetic micro-structures were constructed with the open-source 3D polycrystal generator tool Neper [11].

A Python script controlling the open-source CAD software Salome [12] introduces the grain boundary phase with a specific thickness and produces the finite element mesh. For these synthetic microstructures the demagnetization curve is computed through minimization of the micromagnetic energy with a preconditioned nonlinear conjugate gradient method [13]. The search for higher coercive fields, $\mu_0 H_c$, and energy density products, $(BH)_{max}$, is managed via the open-source optimization framework Dakota [14]. Thus, the optimal structure for a given hard-magnetic phase can be found. For characterizing the magnet, we use the $M(H_{ext})$-loop, which gives the magnetization as function of the external field. The demagnetization curve is then corrected by the demagnetizing field of the sample. A similar procedure is done in experiments when the hysteresis curves are not measured in a closed circuit. Then we transform the magnetization to the magnetic induction, $B$, in order to obtain the $B(H_{int})$-loop and the energy density product. Here $H_{int}$ is the internal field.

Finally, we consider the reduction of coercivity by thermal activation. We compute the critical value of the external field that reduces the energy barrier for nucleation to 25 $k_B T$. The system is assumed to overcome this energy barrier within a waiting time of one second owing to thermal fluctuations [15]. We use a modified string method [16] to compute the energy barriers for different values of the external field. The computation of reduction of the coercive field through thermal activation gives the limits of the coercive field [4] of a certain hard magnetic phase.

## 3 Results

### 3.1 Rare-earth free phases

Using an adaptive genetic algorithm [7] the crystal phase space of Fe-Co-Ta was searched for non-cubic systems with high stability. For CoFe$_6$Ta we performed the two simulations starting from the scratch with 8 and 16 atoms/cell. Various non-cubic stable phases could be identified. Some of the most stable non-cubic phases were tetragonal (space group 115), rhombohedral (space group 160), orthorhombic (space group 38), and orthorhombic (space group 63, where $a$ and $b$ lattice parameters are very similar), and with an enthalpy of formation of -0.07033 eV/atom, -0.06353 eV/atom, -0.06025 eV/atom and -0.05929 eV/atom, respectively. Data and calculations details of these theoretical phases can be found in the Novamag database [17], see the following links [18]. The lowest ground state energy was found for a monoclinic system (space group 8) with enthalpy of formation of -0.07488 eV/atom [19]. These results correspond to a high-throughput DFT calculations (at zero-temperature) using AGA, where similar default settings were used for all of them with the Generalized Gradient Approximation (GGA). To analyze in more detail the stability of these phases is recommended to compute the free energy at finite temperature including electronic, phononic and magnetic terms [20]. In space groups 63 and 160, CoFe$_6$Ta shows a uniaxial magnetocrystalline anisotropy. The complete theoretical study of these phases is in progress and it is planned to be reported in the near future, so here we just selected and mentioned some preliminary results. Using the RSPt code [8] we calculated the anisotropy constant and the spontaneous magnetization for CoFe$_6$Ta in space group 63 to be $K$ = 1 MJ/m³ and $\mu_0 M_s$ = 1.82 T.

Figure 1 shows the micromagnetically computed $B(H_{int})$ loop for different nanostructures made of CoFe$_6$Ta. The grains have approximately the same volume of $34 \times 34 \times 146$ nm³, $56 \times 56 \times 56$ nm³, and $72 \times 72 \times 34$ nm³ for the columns, equiaxed polyhedra, and platelets, respectively. The macroscopic shape of the magnet is cubical with an edge length of 300 nm. The volume fraction of a non-magnetic grain boundary phase is 18 percent. The energy density product is 425 kJ/m³.

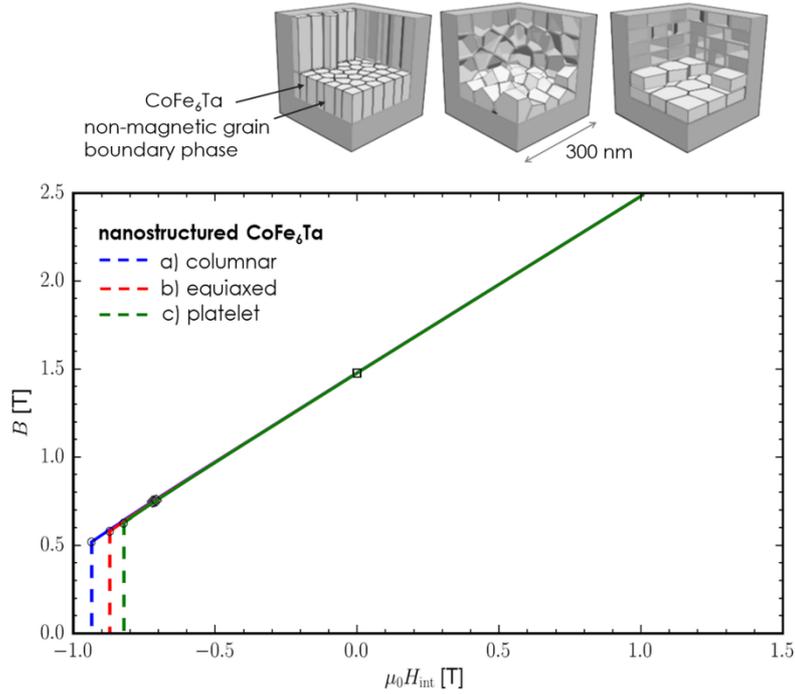

**Figure 1** Magnetic induction as function of the internal field for nanostructured CoFe$_6$Ta. Nanostructuring is essential to obtain a high coercive field. The coercive field increases with increasing aspect ratio of the grains. The aspect ratios of the columnar, equiaxed, and platelet shaped grains 4.3, 1, and 0.47, respectively

Fe-rich materials with non-cubic uniaxial crystal structures are promising candidates for rare-earth free permanent magnets. Because of the hexagonal crystal structure and its high spontaneous magnetization, Fe$_3$Sn compounds were considered. However, Fe$_3$Sn shows an easy-plane anisotropy [21] both in simulations and experiment. Substituting Sn by Sb changes the easy-plane anisotropy to uniaxial anisotropy. The results show a uniaxial anisotropy constant $K = 0.33$ MJ/m³ and spontaneous magnetization $\mu_0 M_s = 1.52$ T for Fe$_3$Sn$_{0.75}$Sb$_{0.25}$ [22]. These properties were assigned to the grains of a synthetically generated structure whereby the average grain size was 50 nm. An exchange stiffness constant $A = 10$ pJ/m was used. The grains were separated by a weakly ferromagnetic grain boundary (gb) phase with a magnetization of $\mu_0 M_{s,gb} = 0.81$ T and an exchange stiffness constant $A_{gb} = 3.7$ pJ/m. The micromagnetic simulation of the reversal process (see Figure 2) shows that multidomain states remain stable after irreversible switching owing to domain wall pinning at the grain boundaries. The computed energy density product is coercivity limited. Its maximum value of 290 kJ/m³ may only be achieved for nanostructured systems with a grain size smaller than 50 nm. Unfortunately, Fe$_3$Sn$_{0.75}$Sb$_{0.25}$ is not stable. Attempts to stabilize the phase by small additions of Mn were successful. However, owing to the change of the electronic structure and the number of valence electrons the anisotropy flipped back to in-plane again [22].

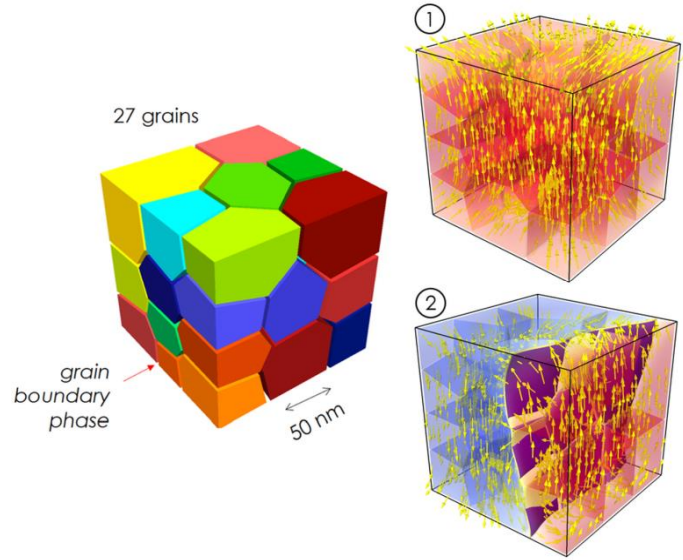

**Figure 2** Domain wall – microstructure interaction in a $Fe_3Sn_{0.75}Sb_{0.25}$ magnet. Left hand side: Grain structure. Right hand side: At an internal field of $\mu_0 H_{int} = 0.49$ T the flower like magnetic state (1) breaks into a two-domain state (2) with a domain wall pinned at the grain boundaries. Images reproduced from [22].

### 3.3 Microstructure optimization

For computing the influence of the microstructure on the hysteresis properties we varied the grain size, the grain shape, the thickness of the grain boundary phase, and the magnetization in the grain boundary phase. The design space was sampled with the help of the software tool Dakota [14].

In order to obtain a general trend on how microstructural features influence the coercive field we use dimensionless units. The coercive field is given in units of the anisotropy field, $2K/(\mu_0 M_s)$. The grain boundary magnetization is measured in units of the magnetization of the main hard magnetic phase, $M_{s,bulk}$. Grain size and grain boundary thickness are measured in units of the Bloch parameter $\delta_0 = (A/K)^{1/2}$, which is the characteristic length in hard magnetic materials. The results presented in the Figure 3 and Figure 4 were obtained by varying the microstructure for magnets made of the $L1_0$ FeNi (bulk), MnAl, and $Nd_{0.2}Zr_{0.8}Fe_{10}Si_2$ (see Table 1). The granular structure used for the simulations is shown in the top row of Figure 3. Because we used dimensional units, the influence of grain boundary phase, grain size, and grain aspect ratio on coercivity for other hard magnetic phases can be derived from the presented data.

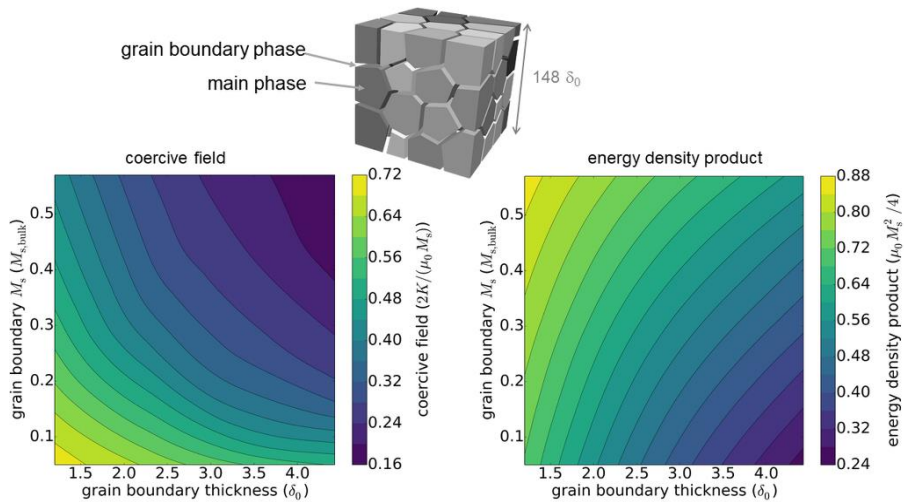

**Figure 3** Coercive field (left hand side) and the energy density product (right hand side) as function of grain boundary properties.

The design space for analysis of the influence of grain boundary properties on coercivity and energy density product was spanned by the grain boundary thickness and the magnetization of the grain boundary. We varied the thickness of the grain boundary from 1.1 $\delta_0$ to 4.4 $\delta_0$, while keeping the size of the magnet constant. The magnetization of the grain boundary phase was varied from 0.05 $M_{s,bulk}$ to 0.55 $M_{s,bulk}$. The exchange stiffness constant of the grain boundary phase is assumed to be proportional to its magnetization squared [23] according to $A_{gb} = A_{bulk}(M_{s,gb}/M_{s,bulk})^2$. Thus, the grain boundary phase changes from almost non-magnetic to ferromagnetic. The polycrystalline structure used for the simulations is shown in Figure 3. The average grain size is 37 $\delta_0$.

Clearly the maximum coercive field is reached for a thin, almost non-magnetic grain boundary phase. Both, increasing the grain boundary thickness or increasing the grain boundary magnetization reduces the coercive field. The magnetization of the grain boundary phase contributes to the total magnetization. Therefore, the maximum energy density product occurs for thin grain boundaries and a moderately high magnetization in the grain boundary. We can conclude that excellent hysteresis properties can be achieved even for ferromagnetic grain boundaries, given that its thickness is sufficiently small. For example, a coercive field of $0.4 \times 2K/(\mu_0 M_s)$ is reached for a grain boundary thickness of 2 $\delta_0$, when the magnetization in the grain boundary phase is about ½ of its bulk value.

The weakly soft magnetic grain boundary phase acts as soft magnetic defect. Detailed micromagnetic studies show that at such grain boundaries magnetization reversal is initiated [24]. We see that the coercive field decreases with increasing spontaneous magnetization of the grain boundary phase. Furthermore, the coercive field decreases with increasing thickness of the grain boundary phase. Though the structure is more complicated for polycrystalline magnets with a weakly soft magnetic grain boundary phase, the effect is similar to that reported by Richter et al. [25] who showed a similar dependence of the nucleation field on the size of a soft defect in a one-dimensional micromagnetic model. The energy to form the domain wall of the reversed nucleus increases with decreasing thickness of the soft magnetic defect. In magnets with a thin grain boundary phase the domain wall of the nucleus extends into the main hard magnetic phase and the domain wall energy increases. Therefore, magnets with a thinner grain boundary phase show a higher coercive field.

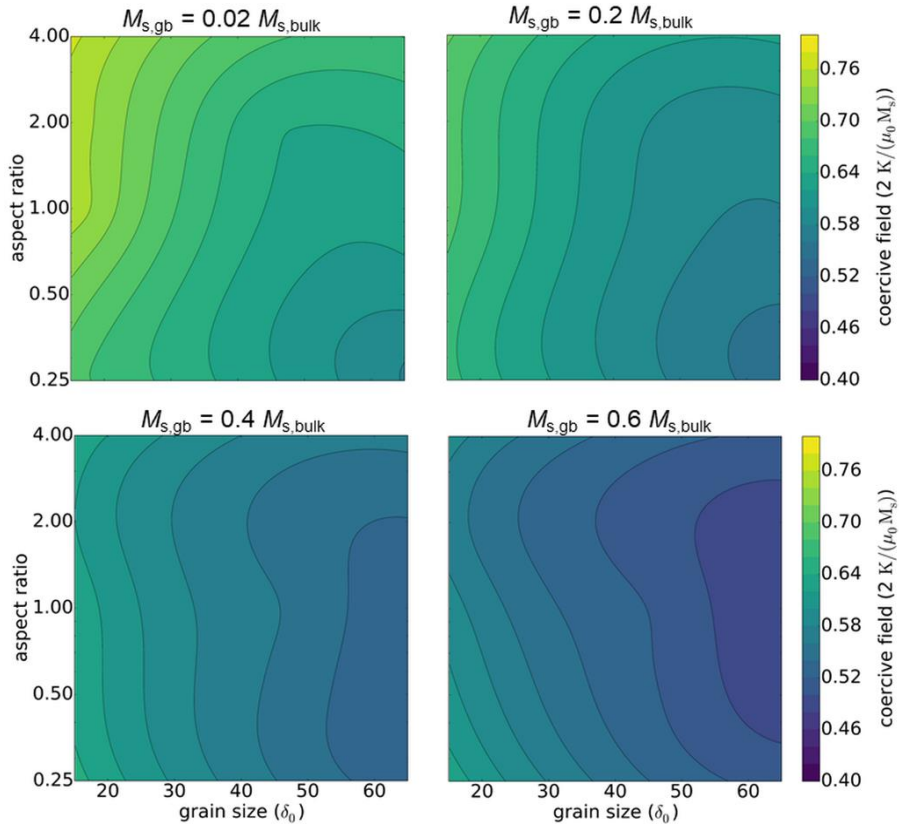

**Figure 4** Influence of grain size and grain shape. The contours give the coercive field as function of grain size and aspect ratio. The different panels refer to different saturation magnetization of the grain boundary phase with a thickness of $\delta_0$.

We now modified the design space. We kept the grain boundary thickness at $\delta_0$ and varied the magnetization in the grain boundary phase, the size of the grains, and the aspect ratio of the grains. An aspect ratio greater 1 refers to elongated, needle like grains; an aspect ratio smaller 1 refers to platelet like grains.

The panels of Figure 4 show the coercive field as function of the grain size and the aspect ratio for different magnetization in the grain boundary phase. For an almost non-magnetic grain boundary phase the coercive field increases with increasing aspect ratio. This means that magnets with needle like grains show a higher coercive field than magnets with platelet like grains. This effect diminishes when the magnetization of the grain boundary phase is increased. For $M_{s,gb} = 0.4\ M_{s,bulk}$, there is hardly any change of the coercive field with aspect ratio. For large magnetization in the grain boundary phase the trend is reversed and platelet-shaped grains show a slightly higher coercive field than needle like grains. The grain size effect on coercivity is more pronounced in platelet shaped grains.

The results of Figure 3 also show that the highest coercive field can be achieved for an almost non-magnetic grain boundary (0.05 $M_{s,bulk}$). The coercive field is a factor of 4.5 higher than for a grain boundary phase with a spontaneous magnetization of 0.55 $M_{s,bulk}$. Figure 4 shows that the coercivity increases with decreasing grain size. We can conclude that magnets with small, exchange-decoupled grains show the highest coercive field. Indeed, the highest coercive field is found for the top left point on the top left subplot for Figure 4 with $M_{s,gb} = 0.02\ M_{s,bulk}$: Here we have a nanostructured systems with exchange isolated grains with a grain size which is smaller than 20 $\delta_0$.

### 3.4 Coercivity limits

Using numerical micromagnetics we can separate the effects that lead to a reduction of coercivity with respect to the anisotropy field of the magnet. We compute the demagnetization curve but switch off the magnetostatic field. When the computed coercive field is less than the anisotropy field the reduction has to be attributed to misalignment of the grains or secondary soft magnetic phases. In a second step, we switch on magnetostatic interactions and simulate the demagnetization curve again. The resulting decrease of the coercive field has to be attributed to demagnetizing effects. Finally, we can simulate how the system escapes from a metastable state over the lowest energy barrier. This gives the temperature dependent coercive field [4].

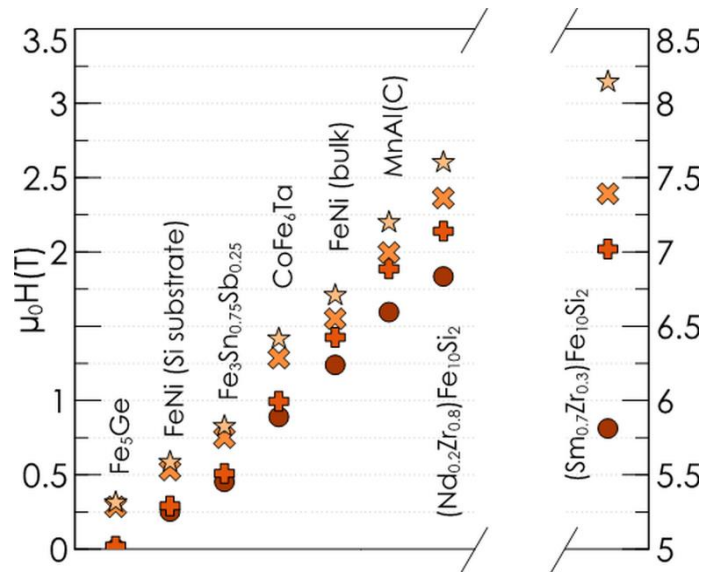

**Figure 5** Limits of coercivity. Effects that reduce the coercive field in permanent magnets for different candidate phases. The symbols give the coercive field. **stars**: anisotropy field, **x**: micromagnetics without magnetostatics, **+**: full micromagnetics, **o**: micromagnetics with thermal activation. Please note the different scale for the $\mu_0 H$ axis for $(Sm,Zr)Fe_{10}Si_2$.

In the following analysis we did not assume any soft magnetic secondary phases. The external field was oriented one degree off the easy axes of a small cube with an edge length of 40 nm. The computed effects that reduce the anisotropy fields are (1) misorientation, (2) demagnetizing effects, and (3) thermal fluctuations. Here the coercive field was computed for an ideal structure: The grain size is very small (40 nm) and there are no defects. Thus, the computed coercive field is an upper limit for coercivity for a given hard magnetic phase.

We applied this procedure to several candidate phases for rare-earth free or rare earth reduced magnets. For each phase we show the anisotropy field, the reduction owing to misorientation, the reduction by demagnetizing effects, and the reduction by thermal fluctuations (see Figure 5). The intrinsic magnetic properties used for the simulations are listed in Table 1. The anisotropy constant, the spontaneous magnetization, and the exchange constant for MnAl were obtained from atomistic spin dynamics at $T = 300$ K. $Fe_5Ge$ is an Fe-rich binary phase predicted by an adaptive genetic algorithm. The anisotropy constant and the spontaneous magnetization for $Fe_5Ge$, $Fe_3Sn_{0.75}Sb_{0.25}$, and $CoFe_6Ta$ were obtained from first principle simulations at $T = 0$. The exchange constant for $Fe_5Ge$ and $CoFe_6Ta$ was taken to be proportional to the spontaneous magnetization squared ($A = cM_s^2$) whereby $c$ was taken from $M_s$ and $A$ of α-Fe. The intrinsic material parameters for $L1_0$ FeNi, $Nd_{0.2}Zr_{0.8}Fe_{10}Si_2$, and $Sm_{0.7}Zr_{0.3}Fe_{10}Si_2$ are experimental data for $T = 300$ K were taken from literature. If no other source for the value of the exchange constant was available we used $A = 10$ pJ/m [26].

The results clearly show that we cannot expect a coercive field greater than 1 T in most rare earth free magnets. For FeNi (bulk) a high degree of uniform chemical order was assumed. Experimentally synthesized $L1_0$ FeNi particles may contain patches where the chemical order is reduced locally. The corresponding local reduction of magnetocrystalline anistropy will deteriorate coercivity. Similarly, crystal defects such as twins or antiphase boundaries reduce the coercive field in MnAl magnets [27]. Rare-earth magnets in the $ThMn_{12}$ structure with Zr substitution have a low rare-earth content. Moreover, the magnetocrystalline anisotropy – especially that of the $(Sm,Zr)Fe_{10}Si_2$ magnet – is sufficiently high to support a reasonable coercive field. For $Nd_{0.2}Zr_{0.8}Fe_{10}Si_2$, and $Sm_{0.7}Zr_{0.3}Fe_{10}Si_2$ the coercive field computed with thermal activation (dots in Figure 5) is 70 percent of the anisotropy field.

**Table 1** Anisotropy constant $K$, spontaneous magnetization Ms, and exchange constant $A$ used for the simulations presented in Figure 5.

| Phase | $K$ (MJ/m³) | $\mu_0 M_s$ (T) | $A$ (pJ/m) | |
|---|---|---|---|---|
| $Fe_5Ge$ | 0.23 | 1.8 | 14.7 | |
| $L1_0$ FeNi (Si substrate) | 0.38 | 1.5 | 10 | [28] |
| $Fe_3Sn_{0.75}Sb_{0.25}$ | 0.33 | 1.52 | 10 | [22] |
| $CoFe_6Ta$ | 1 | 1.82 | 14.9 | |
| $L1_0$ FeNi (bulk) | 1.1 | 1.38 | 10 | [29] |
| MnAl | 0.7 | 0.8 | 7.6 | [30] |
| $Nd_{0.2}Zr_{0.8}Fe_{10}Si_2$ | 1.16 | 1.12 | 10 | [31] |
| $Sm_{0.7}Zr_{0.3}Fe10Si_2$ | 3.5 | 1.08 | 10 | [32] |

## 4 Conclusions

We showed how to exploit materials simulations for the computational design of the next generation rare-earth reduced permanent magnets. Based on the results presented above we can draw the following conclusions.

- Nanostructuring is essential to achieve a high coercive field in rare-earth free compounds with moderate magneto-crystalline anisotropy.
- Coercivity decreases with increasing magnetization in the grain boundary phase and with increasing thickness of the grain boundary phase.
- However, excellent permanent magnetic properties can be achieved even for moderately ferromagnetic grain boundary phases provided that the grain boundary is thin enough.
- The shape of the grains is only important for nearly non-magnetic grain boundaries. For systems in which ferromagnetic Fe containing grain boundaries are expected, the grain shape plays a minor role.
- Thermal fluctuations may considerably reduce the coercive field. Thus, even in perfect structures the coercive field is well below the anisotropy field.

## Acknowledgements

This work was supported by the EU H2020 project NOVAMAG (Grant no 686056), the Austrian Science Fund FWF (I3288-N36), and by the European Regional Development Fund in the IT4Innovations national supercomputing center - path to exascale project, project number CZ.02.1.01/0.0/0.0/16_013/0001791 within the Operational Programme Research, Development and Education.

## Compliance with ethics guidelines

All authors declare that they have no conflict of interest or financial conflicts to disclose.

## References

[1] Constantinides S. Magn Mag 2016;Spring 2016:6.
[2] Nakamura H. The current and future status of rare earth permanent magnets. Scr Mater 2017.
[3] Coey JMD. Permanent magnets: Plugging the gap. Scr Mater 2012;67:524–9. doi:10.1016/j.scriptamat.2012.04.036.
[4] Fischbacher J, Kovacs A, Oezelt H, Gusenbauer M, Schrefl T, Exl L, et al. On the limits of coercivity in permanent magnets. Appl Phys Lett 2017;111:072404. doi:10.1063/1.4999315.
[5] Nieves P, Arapan S, Hadjipanayis GC, Niarchos D, Barandiaran JM, Cuesta-López S. Applying high-throughput computational techniques for discovering next-generation of permanent magnets: Applying high-throughput computational techniques for discovering next-generation of permanent magnets. Phys Status Solidi C 2016;13:942–50. doi:10.1002/pssc.201600103.
[6] Nieves P, Arapan S, Cuesta-Lopez S. Exploring the Crystal Structure Space of $CoFe_2$ P by Using Adaptive Genetic Algorithm Methods. IEEE Trans Magn 2017;53:1–5. doi:10.1109/TMAG.2017.2727880.
[7] Arapan S, Nieves P, Cuesta-López S. A high-throughput exploration of magnetic materials by using structure predicting methods. J Appl Phys 2018;123:083904.
[8] Wills JM, Alouani M, Andersson P, Delin A, Eriksson O, Grechnyev O. Full-Potential Electronic Structure Method: energy and force calculations with density functional and dynamical mean field theory. vol. 167. Springer Science & Business Media; 2010.
[9] Fischbacher J, Kovacs A, Gusenbauer M, Oezelt H, Exl L, Bance S, et al. Micromagnetics of rare-earth efficient permanent magnets. J Phys Appl Phys 2018;51:193002. doi:10.1088/1361-6463/aab7d1.

[10] Kresse G, Joubert D. From ultrasoft pseudopotentials to the projector augmented-wave method. Phys Rev B 1999;59:1758.
[11] Quey R, Renversade L. Optimal polyhedral description of 3D polycrystals: method and application to statistical and synchrotron X-ray diffraction data. Comput Methods Appl Mech Eng 2018;330:308–333.
[12] Salome. http://www.salome-platform.org/ (accessed February 1, 2018).
[13] Exl L, Fischbacher J, Kovacs A, Oezelt H, Gusenbauer M, Schrefl T. Preconditioned nonlinear conjugate gradient method for micromagnetic energy minimization. Comput Phys Commun 2019;235:179–186. doi:10.1016/j.cpc.2018.09.004.
[14] Adams BM, Bohnhoff W, Dalbey K, Eddy J, Eldred M, Gay D, et al. DAKOTA, a multilevel parallel object-oriented framework for design opti-mization, parameter estimation, uncertainty quantification, and sensitivity analy-sis: version 5.0 user's manual. Sandia Natl Lab Tech Rep SAND2010-2183 2009.
[15] Gaunt P. Magnetic viscosity in ferromagnets: I. Phenomenological theory. Philos Mag 1976;34:775–80. doi:10.1080/14786437608222049.
[16] Carilli MF, Delaney KT, Fredrickson GH. Truncation-based energy weighting string method for efficiently resolving small energy barriers. J Chem Phys 2015;143:054105. doi:10.1063/1.4927580.
[17] Nieves P et al. Novamag database, to be published, http://crono.ubu.es/novamag/.
[18] http://crono.ubu.es/novamag/show_item_features?mafid=1574, http://crono.ubu.es/novamag/show_item_features?mafid=1545, http://crono.ubu.es/novamag/show_item_features?mafid=1534, http://crono.ubu.es/novamag/show_item_features?mafid=1579.
[19] http://crono.ubu.es/novamag/show_item_features?mafid=1551.
[20] Lizárraga R, Pan F, Bergqvist L, Holmström E, Gercsi Z, Vitos L. First Principles Theory of the hcp-fcc Phase Transition in Cobalt. Sci Rep 2017;7:3778. doi:10.1038/s41598-017-03877-5.
[21] Sales BC, Saparov B, McGuire MA, Singh DJ, Parker DS. Ferromagnetism of Fe3Sn and Alloys. Sci Rep 2015;4:7024. doi:10.1038/srep07024.
[22] Vekilova O Yu, Fayyazi B, Skokov KP, Gutfleisch O, Echevarria-Bonet C, Barandiaran JM, et al. Tuning magnetocrystalline anisotropy of $Fe_3Sn$ by alloying. Phys. Rev. B, in press, ArXiv Prepr ArXiv180308292 2018.
[23] Kronmüller H, Fähnle M. Micromagnetism and the microstructure of ferromagnetic solids. Cambridge University Press; 2003.
[24] Zickler GA, Fidler J, Bernardi J, Schrefl T, Asali A. A Combined TEM/STEM and Micromagnetic Study of the Anisotropic Nature of Grain Boundaries and Coercivity in Nd-Fe-B Magnets. Adv Mater Sci Eng 2017;2017:6412042. doi:10.1155/2017/6412042.
[25] Richter HJ. Model calculations of the angular dependence of the switching field of imperfect ferromagnetic particles with special reference to barium ferrite. J Appl Phys 1989;65:3597–601. doi:10.1063/1.342638.
[26] Wang D, Sellmyer DJ, Panagiotopoulos I, Niarchos D. Magnetic properties of $Nd(Fe,Ti)_{12}$ and $Nd(Fe,Ti)_{12}N_x$ films of perpendicular texture. J Appl Phys 1994;75:6232–4. doi:10.1063/1.355408.
[27] Bance S, Bittner F, Woodcock TG, Schultz L, Schrefl T. Role of twin and anti-phase defects in MnAl permanent magnets. ACTA Mater 2017;131:48–56. doi:10.1016/j.actamat.2017.04.004.
[28] Kovacs A, Ozelt H, Fischbacher J, Schrefl T, Kaidatzis A, Salikhof R, et al. Micromagnetic Simulations for Coercivity Improvement through Nano-Structuring of Rare-Earth Free $L1_0$-FeNi Magnets. IEEE Trans Magn 2017;53:7002205, doi:10.1109/TMAG.2017.2701418.
[29] Niarchos D, Gjoka M, Psycharis V, Devlin E. Towards realization of bulk $L1_0$-FeNi, IEEE; 2017, doi:10.1109/INTMAG.2017.8007560.
[30] Nieves P, Arapan S, Schrefl T, Cuesta-Lopez S. Atomistic spin dynamics simulations of the MnAl τ-phase and its antiphase boundary. Phys Rev B 2017;96:224411. doi:10.1103/PhysRevB.96.224411.
[31] Gjoka M, Psycharis V, Devlin E, Niarchos D, Hadjipanayis G. Effect of Zr substitution on the structural and magnetic properties of the series $Nd_{1-x}Zr_xFe_{10}Si_2$ with the $ThMn_{12}$ type structure. J Alloys Compd 2016;687:240–5. doi:10.1016/j.jallcom.2016.06.098.
[32] Gabay AM, Cabassi R, Fabbrici S, Albertini F, Hadjipanayis GC. Structure and permanent magnet properties of $Zr_{1-x}R_xFe_{10}Si_2$ alloys with R = Y, La, Ce, Pr and Sm. J Alloys Compd 2016;683:271–5. doi:10.1016/j.jallcom.2016.05.092.